\documentclass[showpacs,preprintnumbers,amsmath,amssymb]{revtex4-1}
\usepackage[latin9]{inputenc}
\usepackage{amsmath}
\usepackage{amssymb}
\usepackage{graphicx}
\usepackage[unicode=true,
 bookmarks=true,bookmarksnumbered=false,bookmarksopen=false,
 breaklinks=false,pdfborder={0 0 1},backref=false,colorlinks=false]
 {hyperref}
\begin{document}
\title{Semi-transparent boundaries in CPT-even Lorentz violating electrodynamics}
\author{L. H. C. Borges}
\email{luizhenriqueunifei@yahoo.com.br}

\author{A. F. Ferrari}
\email{alysson.ferrari@ufabc.edu.br}

\affiliation{Universidade Federal do ABC - UFABC, Centro de Ciências Naturais e
Humanas, Rua Santa Adélia, 166, 09210-170, Santo André, SP, Brazil}
\begin{abstract}
Some aspects of the nonbirefringent CPT-even gauge sector of the Standard
Model Extension (SME), in the vicinity of a semi-transparent mirror,
are investigated in this paper. We first consider a model where the
Lorentz symmetry breaking is caused by a single background vector
$v^{\mu}$, and we obtain perturbative results up to second order
in $v^{\mu}$. Specifically, we compute the modified propagator for
the gauge field due to the presence of the mirror and we analyze the
corresponding interaction between the mirror and a stationary point-like
charge. We show that when the charge is placed in the vicinity of
the mirror, a spontaneous torque emerges, which is a new effect with
no counterpart in Maxwell electrodynamics. We also compare these results
with the corresponding ones obtained for the Lorentz violating scalar
field theory. As expected, in the limiting case of perfect mirrors,
we recover the interaction found via the image method. Finally, we
discuss how we can extend these results for a more general Lorentz
violating background.
\end{abstract}
\maketitle

\section{\label{I}Introduction}

In recent years theories with Lorentz symmetry breaking have been
a subject of intense investigation in the literature, mostly in the
framework of the Standard Model Extension (SME) \citep{SME1,SME2}.
Some aspects of Lorentz symmetry breaking have been studied, for instance,
in classical \citep{LHCFABJHN,fontes2,fontes3,CED1,CED2,CED3,CED4,CED5,CED6,CED7,CED8,LHCAFFcptodd}
and quantum \citep{QED1,QED2,QED3,QED4,QED5,QED6,QED7,QED8} electrodynamics,
radiative corrections \citep{RC1,RC2,RC3,RC4,RC5,RC6,RC7,RC8,RC9},
topological defects \citep{TPD1,TPD2,TPD3,TPD4}, electromagnetic
wave propagation \citep{wave1,wave2}, gravity theories \citep{G1,G2,G3,G4,G5,G6,G7},
noncommutative field theories \citep{NC1,NC2,NC3}, among others.
On the other hand, the study of models in the presence of nontrivial
boundary conditions is of great interest, with a large number of
applications in several branches of physics. We can cite, for example,
the use of $\delta$-like potentials coupled to quantum fields describing
semi-transparent mirrors in order to study the Casimir effect \citep{Milton,BorUM,KimballA,BordKD,NRVMH,NRVMMH2,PsRj,Caval,FABFEB2},
the calculation of the interaction energy between point-like field
sources and $\delta$-like mirrors \citep{FABFEB,GTFABFEB,OliveiraBorgesAFF},
as well as studies related to both Lee-Wick and Maxwell-Chern-Simons
electrodynamics in the presence of boundary conditions \citep{FABAAN1,LW1,LW2,LW3,LHCBFEBHLO,BorgesBarone22}.
A topic we believe still deserves more attentionare Lorentz violating
theories in the presence of boundary conditions, since it could be
of great interest to investigate the physical phenomena that can arise
in this scenario. In this context, we can cite some works concerning
the Casimir effect \citep{CSE1,CSE2,CSE3,CSE4,CSE5,CSE6,CSE7,CSE8,CSE9,CSE10,CSE11,CSE12,CSE13,CSE14,CSE15},
the study of Lorentz violating Maxwell electrodynamics in the vicinity
of a perfect conductor \citep{LHCBFABplate,LHCBFABplate2}, and effects
related to the presence of a semi-transparent mirror in a Lorentz
violating scalar field theory \citep{LHCBAFFFAB}. However, the Maxwell
electrodynamics with Lorentz symmetry breaking in the presence of
a semi-transparent mirror have not been considered up to now in the
literature. This is an interesting topic since, in practice, electromagnetic
configurations in actual experiments do usually involve conductors,
which have to be properly considered in the theoretical models. Another
interesting question that can be raised in this scenario concerns
the modifications the gauge propagator undergoes due to the presence
of a single semi-transparent mirror, and the influence of the mirror
on stationary point-like field sources.

This paper is devoted to this subject in the context of the nonbirefringent
CPT-even pure-photon sector of the SME, where we search for Lorentz
violation effects due to the presence of a single semi-transparent
mirror. We consider first the model studied in \citep{LHCFABJHN,LHCBFABplate,LHCFAB},
where the Lorentz violation is caused by a single background vector
$v^{\mu}$. Since the Lorentz breaking parameter should be very small
we treat it perturbatively up second order, which is the lowest order
in which it appears in our calculations. In section \ref{II} we compute
the modification undergone by the Lorentz violating gauge propagator
due to the presence of the mirror. This propagator is used in section
\ref{III} to obtain the interaction energy as well as the interaction
force between a static point-like charge and the mirror. We show that
when the charge is placed in the vicinity of the mirror, a spontaneous
torque emerges in the system, which is a new effect with no counterpart
in Maxwell electrodynamics in the presence of a semi-transparent mirror.
We also compare the obtained results with the similar ones computed
in\,\citep{LHCBAFFFAB} for the Lorentz violating scalar field theory.
As expected, in the limiting case of a perfect mirror, we recover
the results of \,\citep{LHCBFABplate} obtained via image method.
Then, in Section\,\eqref{sec:Some-results-for}, our results are
extended to a more general LV setting by using the image method, together
with a careful consideration of the limiting case described in Sec.\,\ref{III}.
We can then estimate the physical effects in the kind of system considered
by us for a quite general LV background. Section \ref{IV} is devoted
to conclusions and final remarks.

Throughout the paper we work in a $3+1$-dimensional Minkowski space-time
with metric $\eta^{\rho\nu}=(1,-1,-1,-1)$. The Levi-Civita tensor
is denoted by $\epsilon^{\rho\nu\alpha\beta}$ with $\epsilon^{0123}=1$.

\section{\label{II} The propagator in the presence of a semi-transparent
boundary}

In this section, we start by considering the Lagrangian of the CPT-even
photon sector of the minimal SME with the inclusion of a $\delta$-like
term, representing the presence of a semi-transparent boundary or
a two-dimensional semi-transparent mirror. Without loss of generality
and for simplicity, we will consider the mirror to be perpendicular
to the $x^{3}$ axis, at $x^{3}=a$. The corresponding model is given
by 
\begin{equation}
{\cal {L}}=-\frac{1}{4}F_{\mu\nu}F^{\mu\nu}-\frac{1}{2\gamma}\left(\partial_{\mu}A^{\mu}\right)^{2}-\frac{1}{4}\left(K_{F}\right)^{\mu\nu\alpha\beta}F_{\mu\nu}F_{\alpha\beta}-\frac{1}{m}\left(\frac{1}{2}S^{\mu}\epsilon_{\mu\nu\alpha\beta}F^{\alpha\beta}\right)^{2}\delta\left(x^{3}-a\right)-J^{\mu}A_{\mu}\ ,\label{LVMirrorSME}
\end{equation}
where $A^{\mu}$ is the gauge field, $F^{\mu\nu}=\partial^{\mu}A^{\nu}-\partial^{\nu}A^{\mu}$
is the field strength, $\gamma$ is a gauge fixing parameter, $j^{\mu}$
is an external source, $S^{\gamma}=\eta_{\ 3}^{\gamma}$ is the vector
normal to the mirror and $m^{-1}>0$ is a coupling constant with inverse
of mass dimension, establishing the degree of transparency of the
mirror, the limit $m\rightarrow0$ corresponding to a perfect mirror
\citep{FABFEB}. The background tensor $\left(K_{F}\right)^{\mu\nu\alpha\beta}$
is a dimensionless constant having the same symmetries as the Riemann's
tensor and a null double trace $\left(K_{F}\right)_{\ \ \mu\nu}^{\mu\nu}=0$,
which leads to 19 independent components, ten of those being birefringent
and nine nonbirefringent ones. Both for the sake of simplicity, and
due to the fact that the later might be more difficult to constrain
in experiments, due to the lack of the very characteristic effect
of vacuum birefringence, in this paper we are interested in these
nine nonbirefringent components, which are described by the symmetrical
and traceless tensor $k^{\mu\nu}$, related to $\left(K_{F}\right)$
by means of\citep{BAlt}, 
\begin{equation}
\left(K_{F}\right)^{\mu\nu\alpha\beta}=\frac{1}{2}\left(\eta^{\mu\alpha}k^{\nu\beta}-\eta^{\mu\beta}k^{\nu\alpha}+\eta^{\nu\beta}k^{\mu\alpha}-\eta^{\nu\alpha}k^{\mu\beta}\right)\ .\label{paramebire}
\end{equation}
We will consider a particular choice for $k^{\mu\nu}$, in the same
way as in \citep{LHCFABJHN,LHCFAB,LHCBFABplate}, namely 
\begin{equation}
k^{\mu\nu}=v^{\mu}v^{\nu}\thinspace,\label{paramebire2}
\end{equation}
with $v^{2}=v^{\mu}v_{\mu}=0$ in order to ensure the tracelessness
condition. This parametrization does not describe all the nonbirefringent
components of $\left(K_{F}\right)^{\mu\nu\alpha\beta}$, but only
three of the nine independent components described in \eqref{paramebire}.
This choice is made so that we are able to perform analytic calculations
of the various quantities we are interested in. We will discuss the
possibilities of extending our results to more general cases in Sec.\,\eqref{sec:Some-results-for}.
Finally, since $v^{\mu}$ is assumedly very small, along this paper
we treat it perturbatively up to second order, which is the lowest
order in which it contributes to the propagator.

As a result, the model we will consider in this work is described
by 
\begin{equation}
{\cal {L}}=-\frac{1}{4}F_{\mu\nu}F^{\mu\nu}-\frac{1}{2\gamma}\left(\partial_{\mu}A^{\mu}\right)^{2}-\frac{1}{2}v^{\mu}v_{\nu}F_{\mu\lambda}F^{\nu\lambda}-\frac{1}{m}\left(\frac{1}{2}S^{\mu}\epsilon_{\mu\nu\alpha\beta}F^{\alpha\beta}\right)^{2}\delta\left(x^{3}-a\right)-J^{\mu}A_{\mu}\ ,\label{LVMirror}
\end{equation}
 which can be rewritten in the form 
\begin{equation}
{\cal {L}}=\frac{1}{2}A_{\mu}{\cal {O}}^{\mu\nu}A_{\nu}-J^{\mu}A_{\mu}\ ,\label{model1}
\end{equation}
where the differential operator ${\cal {O}}^{\mu\nu}$ is conveniently
separated in two parts, one corresponding to the free theory (without
the mirror) and the other corresponding to the $\delta$-like term,
as follows,
\begin{equation}
{\cal {O}}^{\mu\nu}={\cal {O}}^{(0)\mu\nu}+\Delta{\cal {O}}^{\mu\nu}\ ,\label{model2}
\end{equation}
with the definitions,
\begin{align}
{\cal {O}}^{(0)\mu\nu} & =\left[\Box+(v\cdot\partial)^{2}\right]\eta^{\mu\nu}-\left(1-\frac{1}{\gamma}\right)\partial^{\mu}\partial^{\nu}+v^{\mu}v^{\nu}\Box-\left(v\cdot\partial\right)\left(v^{\mu}\partial^{\nu}+v^{\nu}\partial^{\mu}\right)\ ,\label{operator1}\\
\Delta{\cal {O}}^{\mu\nu} & =\frac{2}{m}\delta\left(x^{3}-a\right)\left(\eta_{\parallel}^{\mu\nu}\Box_{\parallel}-\partial_{\parallel}^{\mu}\partial_{\parallel}^{\nu}\right)\ ,\label{operator2}
\end{align}
where $\Box=\partial_{\mu}\partial^{\mu}$, $v\cdot\partial=v^{\mu}\partial_{\mu}$,
\begin{equation}
\eta_{\parallel}^{\mu\nu}=\eta^{\mu\nu}+\eta^{\mu3}\eta^{\nu3}\thinspace,
\end{equation}
and $\Box_{\parallel}=\partial_{\parallel}^{\alpha}\partial_{\parallel\alpha}$.
We also defined
\begin{equation}
\partial_{\parallel}^{\alpha}=\left(\partial^{0},\partial^{1},\partial^{2}\right)\,,
\end{equation}
due to the fact that the derivatives in the $\delta$-like term in
(\ref{LVMirror}) are taken only in the spacial directions parallel
to the mirror, because of the fixed index in the Levi-Civita tensor,
\begin{equation}
\left(\frac{1}{2}\eta_{\ 3}^{\mu}\epsilon_{\mu\nu\alpha\beta}F^{\alpha\beta}\right)^{2}=\epsilon_{3\alpha\beta\nu}\ \epsilon_{3\rho\tau}^{\ \ \ \ \nu}\left(\partial_{\parallel}^{\alpha}A^{\beta}\right)\left(\partial_{\parallel}^{\rho}A^{\tau}\right)\ .\label{tensor}
\end{equation}

The free propagator satisfies, ${\cal {O}}^{(0)\mu\nu}(x)G_{\nu\beta}^{(0)}\left(x,y\right)=\eta_{\ \beta}^{\mu}\delta^{4}\left(x-y\right)$,
where in the Feynman gauge $(\gamma=1)$ and up to second order in
$v^{\mu}$, we have \citep{LHCFABJHN,LHCBFABplate} 
\begin{equation}
G_{\mu\nu}^{(0)}\left(x,y\right)=-\int\frac{d^{4}p}{(2\pi)^{4}}\frac{e^{-ip\cdot(x-y)}}{p^{2}}\Biggl[\left(1-\frac{(p\cdot v)^{2}}{p^{2}}\right)\eta_{\mu\nu}-v_{\mu}v_{\nu}+\frac{(p\cdot v)}{p^{2}}(p_{\mu}v_{\nu}+v_{\mu}p_{\nu})\Biggr]\ .\label{prop0}
\end{equation}
As it was shown in \citep{FABFEB,GTFABFEB,FABFEB2,LHCBAFFFAB}, the
propagator $G^{\mu\nu}\left(x,y\right)$ which inverts the operator
(\ref{model2}) can be founded recursively in integral form, as follows,
\begin{equation}
G_{\mu\nu}\left(x,y\right)=G_{\mu\nu}^{(0)}\left(x,y\right)-\int d^{4}z\ G_{\mu\gamma}\left(x,z\right)\Delta{\cal {O}}^{\gamma\sigma}\left(z\right)G_{\sigma\nu}^{(0)}\left(z,y\right)\ ,\label{prop1}
\end{equation}
where ${\cal {O}}^{\mu\nu}(x)G_{\nu\beta}\left(x,y\right)=\eta_{\ \beta}^{\mu}\delta^{4}\left(x-y\right)$.
In order to solve Eq.\,(\ref{prop1}), it is convenient to write
$G_{\mu\nu}\left(x,y\right)$ and $G_{\mu\nu}^{(0)}\left(x,y\right)$
as Fourier transforms in the coordinates parallel to the semi-transparent
mirror, 
\begin{align}
G_{\mu\nu}\left(x,y\right) & =\int\frac{d^{3}p_{\parallel}}{\left(2\pi\right)^{3}}\ {\cal {G}}_{\mu\nu}\left(x^{3},y^{3};p_{\parallel}\right)e^{-ip_{\parallel}\cdot\left(x_{\parallel}-y_{\parallel}\right)}\ ,\label{prop2}\\
G_{\mu\nu}^{(0)}\left(x,y\right) & =\int\frac{d^{3}p_{\parallel}}{\left(2\pi\right)^{3}}\ {\cal {G}}_{\mu\nu}^{(0)}\left(x^{3},y^{3};p_{\parallel}\right)e^{-ip_{\parallel}\cdot\left(x_{\parallel}-y_{\parallel}\right)}\ ,\label{prop3}
\end{align}
where $x_{\parallel}^{\mu}=\left(x^{0},x^{1},x^{2}\right)$ and $p_{\parallel}^{\mu}=\left(p^{0},p^{1},p^{2}\right)$
stand for the coordinates and momentum parallel to the mirror, respectively.
We also define ${\cal {G}}_{\mu\nu}\left(x^{3},y^{3};p_{\parallel}\right)$
and ${\cal {G}}_{\mu\nu}^{(0)}\left(x^{3},y^{3};p_{\parallel}\right)$
as being the reduced Green's functions. Splitting Eq.\,(\ref{prop0})
into parallel and perpendicular coordinates with respect to the mirror,
using the fact that \citep{FABAAN1,LHCBFABplate} 
\begin{subequations}
\label{intperp}
\begin{align}
\int\frac{dp^{3}}{2\pi}\frac{e^{ip^{3}(x^{3}-y^{3})}}{p^{2}} & =-\frac{i}{2\Gamma}e^{i\Gamma\mid x^{3}-y^{3}\mid}\label{intperpa}\\
\int\frac{dp^{3}}{2\pi}\frac{e^{ip^{3}(x^{3}-y^{3})}}{p^{4}} & =-\frac{1}{4p_{\parallel}^{2}}\left(\frac{i}{\Gamma}+\mid x^{3}-y^{3}\mid\right)e^{i\Gamma\mid x^{3}-y^{3}\mid}\ ,
\end{align}
\end{subequations}
where $p^{3}$ stands for the momentum perpendicular to the mirror,
$\Gamma=\sqrt{p_{\parallel}^{2}}$, and after performing some manipulations,
the free reduced propagator can be cast as 
\begin{align}
{\cal {G}}_{\mu\nu}^{(0)}\left(x^{3},y^{3};p_{\parallel}\right) & =\frac{i\ e^{i\Gamma\mid x^{3}-y^{3}\mid}}{2\Gamma}\Biggl[f_{1}\left(x^{3},y^{3};p_{\parallel}\right)\eta_{\mu\nu}-v_{\mu}v_{\nu}+\left(p_{\parallel\mu}v_{\nu}+v_{\mu}p_{\parallel\nu}\right)f_{2}\left(x^{3},y^{3};p_{\parallel}\right)\nonumber \\
 & +\left(\eta_{\mu3}v_{\nu}+v_{\mu}\eta_{\nu3}\right)f_{3}\left(x^{3},y^{3};p_{\parallel}\right)\Biggr]\ ,\label{prop4}
\end{align}
where we define the functions 
\begin{align}
f_{1}\left(x^{3},y^{3};p_{\parallel}\right) & =1-\frac{i}{2\Gamma}\Biggl[-\left(\frac{i}{\Gamma}+\mid x^{3}-y^{3}\mid\right)\left(p_{\parallel}\cdot v_{\parallel}\right)^{2}-2\Gamma\left(x^{3}-y^{3}\right)v^{3}\left(p_{\parallel}\cdot v_{\parallel}\right)\nonumber \\
 & +i\Gamma\left(v^{3}\right)^{2}\left[i\Gamma\mid x^{3}-y^{3}\mid+1\right]\Biggr]\ ,\label{f1}\\
f_{2}\left(x^{3},y^{3};p_{\parallel}\right) & =-\frac{i}{2\Gamma}\left(\frac{i}{\Gamma}+\mid x^{3}-y^{3}\mid\right)\left[\left(\frac{i}{\Gamma}+\mid x^{3}-y^{3}\mid\right)\left(p_{\parallel}\cdot v_{\parallel}\right)+\Gamma\left(x^{3}-y^{3}\right)v^{3}\right]\ ,\label{f2}\\
f_{3}\left(x^{3},y^{3};p_{\parallel}\right) & =-\frac{i}{2}\left(x^{3}-y^{3}\right)\left[\left(\frac{i}{\Gamma}+\mid x^{3}-y^{3}\mid\right)\left(p_{\parallel}\cdot v_{\parallel}\right)+\Gamma\left(x^{3}-y^{3}\right)v^{3}\right]\ ,\label{f3}
\end{align}
with $v_{\parallel}^{\mu}=\left(v^{0},v^{1},v^{2}\right)$, and $v^{3}$
standing for the background vector parallel and perpendicular to the
mirror, respectively.

Substituting (\ref{operator2}) into (\ref{prop1}), using (\ref{prop2}),
(\ref{prop3}), (\ref{prop4}), (\ref{f1}), (\ref{f2}) and (\ref{f3}),
after some integrations we arrive at 
\begin{equation}
{\cal {G}}_{\mu\nu}\left(x^{3},y^{3};p_{\parallel}\right)={\cal {G}}_{\mu\nu}^{(0)}\left(x^{3},y^{3};p_{\parallel}\right)+\frac{2p_{\parallel}^{2}}{m}{\cal {G}}_{\mu\gamma}\left(x^{3},a;p_{\parallel}\right)M_{\ \nu}^{\gamma}\left(a,y^{3};p_{\parallel}\right)\ ,\label{prop5}
\end{equation}
with
\begin{align}
M_{\mu\nu}\left(x^{3},y^{3};p_{\parallel}\right) & =\frac{i\ e^{i\Gamma\mid x^{3}-y^{3}\mid}}{2\Gamma}\Biggl[f_{1}\left(x^{3},y^{3};p_{\parallel}\right)\eta_{\parallel\mu\nu}-\left[f_{1}\left(x^{3},y^{3};p_{\parallel}\right)+f_{2}\left(x^{3},y^{3};p_{\parallel}\right)\left(p_{\parallel}\cdot v_{\parallel}\right)\right]\frac{p_{\parallel\mu}p_{\parallel\nu}}{p_{\parallel}^{2}}\nonumber \\
 & -v_{\parallel\mu}v_{\parallel\nu}+\frac{\left(p_{\parallel}\cdot v_{\parallel}\right)}{p_{\parallel}^{2}}p_{\parallel\mu}v_{\parallel\nu}+f_{2}\left(x^{3},y^{3};p_{\parallel}\right)v_{\parallel\mu}p_{\parallel\nu}\nonumber \\
 & +\left[f_{3}\left(x^{3},y^{3};p_{\parallel}\right)-v^{3}\right]\left(v_{\parallel\mu}\eta_{\nu3}-\frac{\left(p_{\parallel}\cdot v_{\parallel}\right)}{p_{\parallel}^{2}}p_{\parallel\mu}\eta_{\nu3}\right)\Biggr]\ .\label{F}
\end{align}

The propagator in (\ref{prop5}) is still defined recursively, but
it is possible to solve for it by setting $y^{3}=a$, thus obtaining
\begin{equation}
{\cal {G}}_{\mu\gamma}\left(x^{3},a;p_{\parallel}\right)\left[\eta_{\ \nu}^{\gamma}-\frac{2p_{\parallel}^{2}}{m}M_{\ \nu}^{\gamma}\left(a,a;p_{\parallel}\right)\right]={\cal {G}}_{\mu\nu}^{(0)}\left(x^{3},a;p_{\parallel}\right)\ ,\label{prop6}
\end{equation}
where ${\cal {G}}_{\mu\nu}^{(0)}\left(x^{3},a;p_{\parallel}\right)$
and $M_{\ \nu}^{\gamma}\left(a,a;p_{\parallel}\right)$ can be obtained
from Eqs.\,(\ref{prop4}) and\,(\ref{F}), respectively. Now, multiplying
both sides of (\ref{prop6}) by the operator that inverts the term
between brackets, we obtain 
\begin{align}
{\cal {G}}_{\mu\nu}\left(x^{3},a;p_{\parallel}\right) & =\frac{i}{2}\left(1-\frac{m}{i\Gamma}\right)^{-1}\frac{e^{i\Gamma\mid x^{3}-a\mid}}{\Gamma}\Biggl\{-\frac{m}{i\Gamma}\left[f_{1}\left(x^{3},a;p_{\parallel}\right)-g_{1}\left(p_{\parallel}\right)\right]\eta_{\parallel\mu\nu}+\Bigl[f_{1}\left(x^{3},a;p_{\parallel}\right)\nonumber \\
 & +f_{2}\left(x^{3},a;p_{\parallel}\right)\left(p_{\parallel}\cdot v_{\parallel}\right)+g_{2}\left(p_{\parallel}\right)\Bigr]\frac{p_{\parallel\mu}p_{\parallel\nu}}{p_{\parallel}^{2}}+\frac{m}{i\Gamma}\left[1-\left(1-\frac{m}{i\Gamma}\right)^{-1}\right]v_{\parallel\mu}v_{\parallel\nu}\nonumber \\
 & -\Biggl[\frac{m}{i\Gamma}f_{2}\left(x^{3},a;p_{\parallel}\right)+\left[1-\left(1-\frac{m}{i\Gamma}\right)^{-1}\right]\frac{\left(p_{\parallel}\cdot v_{\parallel}\right)}{p_{\parallel}^{2}}\Biggr]p_{\parallel\mu}v_{\parallel\nu}+\Bigl[\left(1-\frac{m}{i\Gamma}\right)f_{2}\left(x^{3},a;p_{\parallel}\right)\nonumber \\
 & +g_{3}\left(p_{\parallel}\right)\Bigr]v_{\parallel\mu}p_{\parallel\nu}-\frac{m}{i\Gamma}\left[f_{3}\left(x^{3},a;p_{\parallel}\right)-v^{3}\right]\eta_{\mu3}v_{\parallel\nu}+\Biggl[\left(1-\frac{m}{i\Gamma}\right)v^{3}f_{2}\left(x^{3},a;p_{\parallel}\right)\nonumber \\
 & +\left[f_{3}\left(x^{3},a;p_{\parallel}\right)-v^{3}\right]\frac{\left(p_{\parallel}\cdot v_{\parallel}\right)}{p_{\parallel}^{2}}\Biggr]\eta_{\mu3}p_{\parallel\nu}-\left[\left(1-\frac{m}{i\Gamma}\right)\left[f_{3}\left(x^{3},a;p_{\parallel}\right)-v^{3}\right]+v^{3}\right]v_{\parallel\mu}\eta_{\nu3}\nonumber \\
 & +\left[\frac{i\Gamma}{m}\frac{\left(p_{\parallel}\cdot v_{\parallel}\right)}{p_{\parallel}^{2}}v^{3}-\left(1-\frac{m}{i\Gamma}\right)v^{3}f_{2}\left(x^{3},a;p_{\parallel}\right)\right]p_{\parallel\mu}\eta_{\nu3}+\left(1-\frac{m}{i\Gamma}\right)\Bigl[f_{1}\left(x^{3},a;p_{\parallel}\right)\nonumber \\
 & -2v^{3}f_{3}\left(x^{3},a;p_{\parallel}\right)+\left(v^{3}\right)^{2}\Bigr]\eta_{\mu3}\eta_{\nu3}\Biggr\}\ ,
\end{align}
where we identified the functions 
\begin{align}
g_{1}\left(p_{\parallel}\right) & =\left(1-\frac{m}{i\Gamma}\right)^{-1}\frac{1}{2}\left[\left(v^{3}\right)^{2}-\frac{\left(p_{\parallel}\cdot v_{\parallel}\right)^{2}}{p_{\parallel}^{2}}\right]\ ,\label{g1}\\
g_{2}\left(p_{\parallel}\right) & =\left[1-\left(1-\frac{m}{i\Gamma}\right)^{-1}\right]\frac{1}{2}\left\{ \left(v^{3}\right)^{2}-\left[1-\frac{i}{\Gamma}\left(1+\frac{2p_{\parallel}^{2}}{m}\right)-\frac{1}{m}\right]\frac{\left(p_{\parallel}\cdot v_{\parallel}\right)^{2}}{p_{\parallel}^{2}}\right\} \ ,\label{g2}\\
g_{3}\left(p_{\parallel}\right) & =\left[1-\left(1-\frac{m}{i\Gamma}\right)^{-1}\left(1-\frac{i}{2\Gamma}+\frac{m}{2p_{\parallel}^{2}}\right)\right]\frac{\left(p_{\parallel}\cdot v_{\parallel}\right)}{p_{\parallel}^{2}}\ .\label{g3}
\end{align}

Considering all these results, the modified propagator due to the
presence of the semi-transparent mirror, up to second order in $v^{\mu}$,
reads 
\begin{equation}
G_{\mu\nu}\left(x,y\right)=G_{\mu\nu}^{(0)}\left(x,y\right)+{\bar{G}}_{\mu\nu}\left(x,y\right)\ ,\label{prop9}
\end{equation}
where
\begin{align}
{\bar{G}}_{\mu\nu}\left(x,y\right) & =-\frac{i}{2}\int\frac{d^{3}p_{\parallel}}{\left(2\pi\right)^{3}}\ e^{-ip_{\parallel}\cdot\left(x_{\parallel}-y_{\parallel}\right)}\left(1-\frac{m}{i\Gamma}\right)^{-1}\Biggl\{\left(\eta_{\parallel\mu\nu}-\frac{p_{\parallel\mu}p_{\parallel\nu}}{p_{\parallel}^{2}}\right)\Bigl[f_{1}\left(x^{3},a;p_{\parallel}\right)+f_{1}\left(a,y^{3};p_{\parallel}\right)\nonumber \\
 & -g_{1}\left(p_{\parallel}\right)-1\Bigr]-\Biggl[\frac{i\Gamma}{m}\left[1-\left(1-\frac{m}{i\Gamma}\right)^{-1}\right]\frac{\left(p_{\parallel}\cdot v_{\parallel}\right)^{2}}{p_{\parallel}^{2}}+\left[f_{2}\left(x^{3},a;p_{\parallel}\right)+f_{2}\left(a,y^{3};p_{\parallel}\right)\right]\left(p_{\parallel}\cdot v_{\parallel}\right)\Biggr]\frac{p_{\parallel\mu}p_{\parallel\nu}}{p_{\parallel}^{2}}\nonumber \\
 & -\left[2-\left(1-\frac{m}{i\Gamma}\right)^{-1}\right]v_{\parallel\mu}v_{\parallel\nu}+\Biggl[\left[1+\frac{i\Gamma}{m}\left(1-\left(1-\frac{m}{i\Gamma}\right)^{-1}\right)\right]\frac{\left(p_{\parallel}\cdot v_{\parallel}\right)}{p_{\parallel}^{2}}+f_{2}\left(x^{3},a;p_{\parallel}\right)\Biggr]p_{\parallel\mu}v_{\parallel\nu}\nonumber \\
 & +\Biggl[\left[1-\left(1-\frac{m}{i\Gamma}\right)^{-1}\right]\frac{\left(p_{\parallel}\cdot v_{\parallel}\right)}{p_{\parallel}^{2}}+f_{2}\left(a,y^{3};p_{\parallel}\right)\Biggr]v_{\parallel\mu}p_{\parallel\nu}+\left[f_{3}\left(a,y^{3};p_{\parallel}\right)-v^{3}\right]\left[v_{\parallel\mu}\eta_{\nu3}-\frac{\left(p_{\parallel}\cdot v_{\parallel}\right)}{p_{\parallel}^{2}}p_{\parallel\mu}\eta_{\nu3}\right]\nonumber \\
 & +\left[f_{3}\left(x^{3},a;p_{\parallel}\right)-v^{3}\right]\left[\eta_{\mu3}v_{\parallel\nu}-\frac{\left(p_{\parallel}\cdot v_{\parallel}\right)}{p_{\parallel}^{2}}\eta_{\mu3}p_{\parallel\nu}\right]\Biggr\}\frac{e^{i\Gamma\left(\mid x^{3}-a\mid+\mid y^{3}-a\mid\right)}}{\Gamma}\ .\label{prop10}
\end{align}

The propagator (\ref{prop9}) is composed of the sum of the free propagator
(\ref{prop0}) with the correction (\ref{prop10}), which accounts
for the presence of the semi-transparent mirror. As an important check
of the consistency of our results we point out that by taking the
limit $v^{\mu}\rightarrow0$ in Eq. (\ref{prop10}) we recover the
standard correction to the propagator for the gauge field in the presence
of a single semi-transparent mirror \citep{FABFEB}. Taking the limit
$m\rightarrow0$ in (\ref{prop10}), we recover the correction to
the propagator (\ref{prop0}) due to the presence of a perfect mirror
as obtained in \citep{LHCBFABplate}.

\section{\label{III} Charge-mirror interaction}

Having obtained the relevant propagator in the previous section, here
we consider the interaction energy between a point-like charge and
the semi-transparent mirror, which is given by \citep{FABFEB,FABAAN1,LHCBFABplate,LHCBFEBHLO}
\begin{equation}
E=\frac{1}{2T}\int d^{4}x\ d^{4}y\ J^{\mu}\left(x\right){\bar{G}}_{\mu\nu}\left(x,y\right)J^{\nu}\left(y\right)\ ,\label{energy}
\end{equation}
where $T$ is a time interval, and it is implicit the limit $T\to\infty$
at the end of the calculations.

With no loss of generality, we choose a point-like charge located
at the position ${\bf b}=\left(0,0,b\right)$, perpendicular to the
mirror. The corresponding external source is given by 
\begin{equation}
J^{\mu}\left(x\right)=q\eta^{\mu0}\delta^{3}\left({\bf x}-{\bf b}\right)\ ,\label{source1}
\end{equation}
where the parameter $q$ is a coupling constant between the field
and the delta function, and in this case it can be interpreted as
being the electric charge.

Substituting Eqs. (\ref{source1}) and (\ref{prop10}) in (\ref{energy}),
carrying out the integrals in the order $d^{3}{\bf x}$, $d^{3}{\bf y}$,
$dx^{0}$, $dp^{0}$, $dy^{0}$ and performing some manipulations,
we arrive at 
\begin{align}
E_{MC} & =-\frac{q^{2}}{16\pi^{2}}\int d^{2}{\bf {p}}_{\parallel}\frac{e^{-2R\sqrt{{\bf {p}}_{\parallel}^{2}}}}{m+\sqrt{{\bf {p}}_{\parallel}^{2}}}\Biggl\{1+\frac{1}{2}\frac{\left({\bf {p}}_{\parallel}\cdot{\bf {v}}_{\parallel}\right)^{2}}{{\bf {p}}_{\parallel}^{2}}\Biggl[\frac{2m+\sqrt{{\bf {p}}_{\parallel}^{2}}}{{m+\sqrt{{\bf {p}}_{\parallel}^{2}}}}+2R\sqrt{{\bf {p}}_{\parallel}^{2}}\Biggr]\nonumber \\
 & +\frac{1}{2}\left(v^{3}\right)^{2}\Biggl[\frac{2m+\sqrt{{\bf {p}}_{\parallel}^{2}}}{{m+\sqrt{{\bf {p}}_{\parallel}^{2}}}}-2R\sqrt{{\bf {p}}_{\parallel}^{2}}\Biggr]-\left(v^{0}\right)^{2}\frac{2m+\sqrt{{\bf {p}}_{\parallel}^{2}}}{{m+\sqrt{{\bf {p}}_{\parallel}^{2}}}}\Biggr\}\ ,
\end{align}
where $R=\mid a-b\mid$ is the distance between the mirror and the
charge. The sub-index $MC$ means that we have the interaction energy
between the mirror and the charge. This result can be simplified by
using polar coordinates and integrating out in the solid angle, 
\begin{equation}
E_{MC}=-\frac{q^{2}}{16\pi}\int_{0}^{\infty}dp\ p\ \frac{e^{-2Rp}}{p+m}\Biggl[2+\frac{{\bf {v}}_{\parallel}^{2}}{2}\left(\frac{p+2m}{p+m}+2Rp\right)+\left(v^{3}\right)^{2}\left(\frac{p+2m}{p+m}-2Rp\right)-2\left(v^{0}\right)^{2}\frac{p+2m}{p+m}\Biggr]\ .\label{energyC2}
\end{equation}
Using the fact that \citep{Gradshteyn} 
\begin{align}
\int_{0}^{\infty}dp\ p\ \frac{e^{-2Rp}}{p+m} & =\frac{1}{2R}\left[1-2mRe^{2mR}Ei\left(1,2mR\right)\right]\ ,\label{int1}\\
\int_{0}^{\infty}dp\ p\ \frac{e^{-2Rp}}{p+m}\left(\frac{p+2m}{p+m}+2Rp\right) & =\frac{2}{R}\left[\frac{1}{2}-mR+2\left(mR\right)^{2}e^{2mR}Ei\left(1,2mR\right)\right]\ ,\label{int2}\\
\int_{0}^{\infty}dp\ p\ \frac{e^{-2Rp}}{p+m}\left(\frac{p+2m}{p+m}-2Rp\right) & =0\ ,\label{int3}\\
\int_{0}^{\infty}dp\ p\left(p+2m\right)\ \frac{e^{-2Rp}}{\left(p+m\right)^{2}} & =\frac{1}{R}\left[\frac{1}{2}-mR+2\left(mR\right)^{2}e^{2mR}Ei\left(1,2mR\right)\right]\ ,\label{int4}
\end{align}
where $Ei\left(u,s\right)$ is the exponential integral function \citep{Arfken}
defined by 
\begin{equation}
Ei(n,s)=\int_{1}^{\infty}\frac{e^{-ts}}{t^{n}}\ dt\ \ \,\ \ \ \Re(s)>0\ ,\ n=0,1,2,\cdots\ ,\label{Ei}
\end{equation}
the interaction energy becomes, 
\begin{equation}
E_{MC}=-\frac{q^{2}}{16\pi R}\Biggl\{1-2mRe^{2mR}Ei\left(1,2mR\right)+\left[{\bf {v}}_{\parallel}^{2}-2\left(v^{0}\right)^{2}\right]\left[\frac{1}{2}-mR+2\left(mR\right)^{2}e^{2mR}Ei\left(1,2mR\right)\right]\Biggr\}\ .\label{energyC3}
\end{equation}

Equation (\ref{energyC3}) is a perturbative result up to lowest nontrivial
order in the background vector for the interaction energy of a point
charge and a semi-transparent mirror mediated by the model (\ref{LVMirror}).
The first and second terms on the right hand side reproduces the result
of the standard (Lorentz invariant) electromagnetic field \citep{FABFEB},
the remaining terms are corrections due to the Lorentz symmetry breaking.
We notice that Eq. (\ref{energyC3}) does not depend explicitly on
the component of the background vector perpendicular to the mirror
$v^{3}$. In the limit $m\rightarrow0$, which corresponds to the
field subjected to boundary conditions imposed by a perfect mirror,
the energy (\ref{energyC3}) reads 
\begin{equation}
\lim_{m\rightarrow0}E_{MC}=-\frac{q^{2}}{16\pi R}\left(1-\left(v^{0}\right)^{2}+\frac{1}{2}{\bf {v}}_{\parallel}^{2}\right)\ .\label{energyC4}
\end{equation}
We notice that Eq. (\ref{energyC4}) is equivalent to the result obtained
in Ref. \citep{LHCBFABplate} using the image method in the limiting
case of a perfect mirror, which shows the consistency of our result.
In the limit $m\rightarrow\infty$ the mirror degree of transparency
goes to zero and the energy (\ref{energyC3}) vanishes, as expected.

The force between the point-like charge and the mirror is given by
\begin{align}
F_{MC}=-\frac{\partial E_{MC}}{\partial R} & =-\frac{q^{2}}{16\pi R^{2}}\Biggl\{1-2mR+\left(2mR\right)^{2}e^{2mR}Ei\left(1,2mR\right)\nonumber \\
 & +\left[{\bf {v}}_{\parallel}^{2}-2\left(v^{0}\right)^{2}\right]\Biggl[\frac{1}{2}+2\left(mR\right)^{2}-\left[2\left(mR\right)^{2}+4\left(mR\right)^{3}\right]e^{2mR}Ei\left(1,2mR\right)\Biggr]\Biggr\}\ ,\label{FC}
\end{align}
which is always negative and therefore has an attractive behavior.

\begin{figure}[!h]
\centering \includegraphics{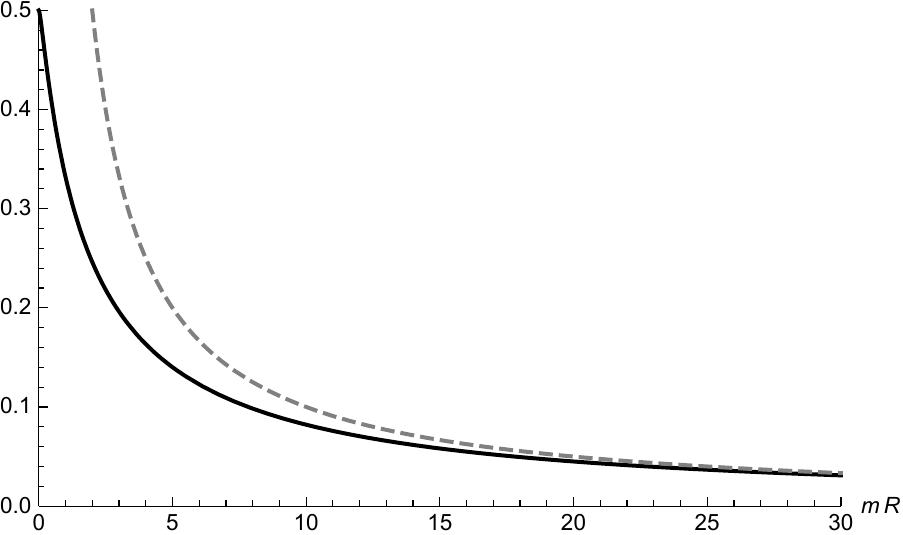}\caption{Plot for the auxiliary function ${\cal {F}}_{1}\left(mR\right)$ (solid
line) defined in Eq. \eqref{fund}. The dashed line is the limiting
function $1/mR$.}
\label{graficocalF|}
\end{figure}

Let us define the following dimensionless function, 
\begin{equation}
{\cal {F}}_{1}\left(x\right)=\frac{1}{2}+2x^{2}-\left(2x^{2}+4x^{3}\right)e^{2x}Ei\left(1,2x\right)\ ,\label{fund}
\end{equation}
and the force (\ref{FC}) becomes 
\begin{equation}
F_{MC}=-\frac{q^{2}}{16\pi R^{2}}\Biggl[1-2mR+\left(2mR\right)^{2}e^{2mR}Ei\left(1,2mR\right)+\left[{\bf {v}}_{\parallel}^{2}-2\left(v^{0}\right)^{2}\right]{\cal {F}}_{1}\left(mR\right)\Biggr]\ ,\label{FC2}
\end{equation}
where we have a Coulombian behavior modulated by the expression inside
brackets. The correction due to the Lorentz symmetry breaking is given
by the function ${\cal {F}}_{1}\left(mR\right)$ which is positive
in its domain as shown in Fig. \ref{graficocalF|}. This function
vanish in the limit $m\rightarrow\infty$, where we have no mirror
present, and its asymptotic behavior for large values of $mR$ is
\begin{equation}
{\cal {F}}_{1}\left(x\right)\underset{x\rightarrow\infty}{\sim}\frac{1}{x}-\frac{9}{4x^{2}}+{\cal O}\left(\frac{1}{x^{3}}\right)\,.
\end{equation}

When we fix the distance between the charge and the mirror, from Eq.
(\ref{energyC3}), we see that the whole system feels a torque depending
on its orientation with respect to the background vector. In order
to calculate this torque, we define as $0\leq\alpha\leq\pi$ the angle
between the normal to the mirror and the background vector ${\bf {v}}$,
in such a way that, ${\bf {v}}_{\parallel}^{2}={\bf {v}}^{2}\sin^{2}\left(\alpha\right)$.
Thus the torque can be computed as follows, 
\begin{equation}
\tau_{MC}=-\frac{\partial E_{MC}}{\partial\alpha}=\frac{q^{2}{\bf {v}}^{2}}{16\pi R}\sin\left(2\alpha\right)\left[\frac{1}{2}-mR+2\left(mR\right)^{2}e^{2mR}Ei\left(1,2mR\right)\right]\ .\label{Torque}
\end{equation}
\begin{figure}[!h]
\centering \includegraphics{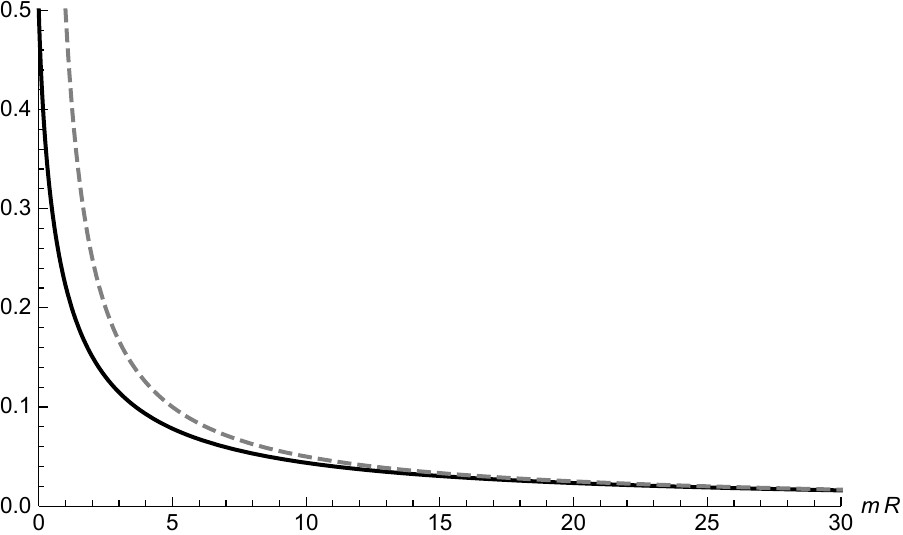}\caption{Plot for the auxiliary function ${\cal {F}}_{2}\left(mR\right)$ defined
in Eq. \eqref{fund2}. The dashed line is the limiting function $1/\left(2mR\right)$.}
\label{graficocalF2|}
\end{figure}

Equation (\ref{Torque}) is a new effect with no counterpart in Maxwell
electrodynamics in the presence of a semi-transparent mirror. For
$\alpha=0,\pi/2,\pi$ the torque vanishes, while for $\alpha=\pi/4$
the torque exhibits a maximum absolute value. Defining the function
\begin{equation}
{\cal {F}}_{2}\left(x\right)=\frac{1}{2}-{\cal Q}\left(x\right)\label{fund2}
\end{equation}
with
\[
{\cal Q}\left(x\right)=x-2x^{2}e^{2x}Ei\left(1,2x\right)\ ,
\]
we can rewrite Eq. (\ref{Torque}) in the form 
\begin{equation}
\tau_{MC}=\frac{q^{2}{\bf {v}}^{2}}{16\pi R}\sin\left(2\alpha\right){\cal {F}}_{2}\left(mR\right)\ .\label{Torque2}
\end{equation}
In Fig. (\ref{graficocalF2|}), we show the behavior of ${\cal {F}}_{2}$
in terms of $mR$. This function is always positive, and goes to zero
if $mR$ is large, with asymptotic behavior, 
\begin{equation}
{\cal {F}}_{2}\left(x\right)\underset{x\rightarrow\infty}{\sim}\frac{1}{2x}-\frac{3}{4x^{2}}+{\cal O}\left(\frac{1}{x^{3}}\right)\,.\label{F2asymptotic}
\end{equation}

It is well-known that the nonbirefringent components $k^{\mu\nu}$
can be divided in parity-even isotropic ${\tilde{\kappa}}_{\mathrm{tr}}$
component, anisotropic parity-even $\left({\tilde{\kappa}}_{e-}\right)^{ij}$
components and the parity-odd $\kappa^{i}$ components, which can
be defined as follows 
\begin{equation}
k^{00}=\frac{3}{2}{\tilde{\kappa}}_{\mathrm{tr}}\ \ ,\ k^{ij}=\frac{1}{2}{\tilde{\kappa}}_{tr}\delta^{ij}-\left({\tilde{\kappa}}_{e-}\right)^{ij}\ \ ,\ k^{0i}=-\kappa^{i}\ .\label{parameters1}
\end{equation}
Now, the obtained results in (\ref{energyC3}), (\ref{FC2}) and (\ref{Torque2})
can be written in terms of the above parameters by using the fact
that, 
\begin{equation}
{\bf {v}}_{\parallel}^{2}-2\left(v^{0}\right)^{2}=-2{\tilde{\kappa}}_{\mathrm{tr}}-\left[\left({\tilde{\kappa}}_{e-}\right)^{11}+\left({\tilde{\kappa}}_{e-}\right)^{22}\right]\ \ ,\ \ {\bf {v}}^{2}=\frac{3}{2}{\tilde{\kappa}}_{\mathrm{tr}}\ .\label{parameters2}
\end{equation}

It is clear from the graphic in Fig. (\ref{graficocalF2|}) and Eq.\,\eqref{F2asymptotic}
that, as $mR$ increases from zero to infinity, ${\cal {F}}_{2}$
decreases monotomically from $1/2$ to $0$, therefore
\begin{equation}
0\leq{\cal Q}\left(mR\right)\leq\frac{1}{2}\ .\label{limitm}
\end{equation}
This fact allows us to make some estimates on the value of ${\cal {F}}_{2}$
and, therefore, the torque, for reasonable values of the parameters
in the theory. We consider a typical distance of atomic experiments
in the vicinity of conductors (mirrors) of order $R\sim10^{-6}$m,
the fundamental electronic charge $q\sim1.60217\times10^{-19}\text{C}$,
and the overestimated value ${\tilde{\kappa}}_{\mathrm{tr}}\sim1.4\times10^{-19}$
obtained from \citep{FRKlin,VANR}. In this case, for a perfect mirror,
corresponding to the limit $m\rightarrow0$, we have ${\cal Q}\left(mR\right)\rightarrow0$
and ${\cal {F}}_{2}\rightarrow1/2$. From Eq\,\eqref{Torque2}, we
obtain a torque of order $\tau_{MC}\sim10^{-41}\,\text{Nm}$. For
an imperfect mirror, the magnitude of the torque is smaller. Taking
$m\sim10^{-5}\text{GeV}$, for example, we obtain $\tau_{MC}\sim10^{-43}\,\text{Nm}$.
This effect is out of reach of being measured by using current technology.

The model (\ref{LVMirror}) can be considered as the electromagnetic
version of the one studied in reference \citep{LHCBAFFFAB}, which
is a Lorentz violating scalar field theory. As discussed in that reference,
even in a Lorentz violation scenario it is always possible to relate
some results obtained for a massless scalar field with the ones obtained
in the corresponding electromagnetic model. For this task, in the
electromagnetic results we must take $v^{0}=0$ and multiply by an
overal factor of $-1$. In a scalar field theory the degree of transparency
of the mirror has mass dimension +1 and therefore an opposite behavior
in relation to the one studied here. However, the connection between
the massless scalar field in the presence of a Dirichlet plane (perfect
mirror) and the gauge field in the presence of a perfect conductor
remains when we have the corresponding Lorentz violating terms in
both theories. This fact can be clarified from Eq. (\ref{energyC4}),
as follows 
\begin{equation}
-\lim_{m\rightarrow0}E_{MC}\left(v^{0}=0\right)=\frac{q^{2}}{16\pi R}\left(1+\frac{1}{2}{\bf {v}}_{\parallel}^{2}\right)\ ,\label{energyC44}
\end{equation}
which is equal to Eq. (96) of \citep{LHCBAFFFAB}. The same connection
can be observed in Lee-Wick electrodynamics \citep{LW3}.

\section{\label{sec:Some-results-for}Some results for more general LV background}

In this section, we make some considerations regarding the generalization
of our results for more general cases described by Eqs.\,\eqref{LVMirrorSME}
and \eqref{paramebire}. We start by defining,
\begin{equation}
f^{\prime}\left(v\right)={\bf {v}}_{\parallel}^{2}-2\left(v^{0}\right)^{2}\ ,\label{fv}
\end{equation}
and rewriting Eq.\,\eqref{FC2} as follows,
\begin{equation}
F_{MC}=-\frac{q^{2}}{16\pi R^{2}}\Biggl[1-2mR+\left(2mR\right)^{2}e^{2mR}Ei\left(1,2mR\right)+f^{\prime}\left(v\right){\cal {F}}_{1}\left(mR\right)\Biggr].\label{FMC1v}
\end{equation}
In the $m\rightarrow0$ limit, we obtain the interaction force between
the charge and the perfect mirror that was found in\,\citep{LHCBFABplate}
via the image method,
\begin{equation}
\lim_{m\rightarrow0}F_{MC}=-\frac{q^{2}}{16\pi R^{2}}\left(1+\frac{1}{2}f^{\prime}\left(v\right)\right)\ .\label{FMC2v}
\end{equation}

In the general case of the model given in Eq.\,\eqref{LVMirrorSME},
we expect to find for the force an expression similar in form, but
with a different function in place of $f^{\prime}$, i.e.,
\begin{equation}
F'_{MC}=-\frac{q^{2}}{16\pi R^{2}}\Biggl[1-2mR+\left(2mR\right)^{2}e^{2mR}Ei\left(1,2mR\right)+f\left(K_{F}\right){\cal {F}}_{1}\left(mR\right)\Biggr],\label{FMC3k}
\end{equation}
where $f\left(K_{F}\right)$ is a function such that, given the particular
choice defined in Eqs.\,\eqref{paramebire} and \eqref{paramebire2},
leads to $f\left(K_{F}\right)\rightarrow f^{\prime}\left(v\right)$.
Similarly, in the $m\rightarrow0$ we should have
\begin{equation}
\lim_{m\rightarrow0}F'_{MC}=-\frac{q^{2}}{16\pi R^{2}}\left(1+\frac{1}{2}f\left(K_{F}\right)\right)\ .\label{FMC4k}
\end{equation}

As we have discussed in\,\citep{LHCBFABplate}, the image method
remains valid in that particular LV setting, and we assume the same
happens here. This is a reasonable assumption, since one of the essential
requirements for the application of the image method is the linearity
of the equations of motion -- which is preserved here -- but not
a completely trivial one, since another requirement is to have a sufficiently
symmetric system. Lorentz violation evidently reduces the symmetry
of the problem. However, the configuration discussed here is similar
enough to the one treated in\,\citep{LHCBFABplate} to give us confidence
in using the image method for our considerations.

Thus, we consider that the result in Eq.\,(\ref{FMC4k}) is equivalent
to the interaction force, up to the first order in $K_{F}$, between
two point charges $q_{1}=q$ and $q_{2}=-q$ separated by a distance
$2R$. This result can be easily obtained from\,\citep{fontes3}
as
\begin{equation}
F_{CC}^{\prime}=-\frac{q^{2}}{16\pi R^{2}}\left[\left(K_{F}\right)^{0101}+\left(K_{F}\right)^{0202}\right],\label{FMC5k}
\end{equation}
where the $CC$ means this is the interaction force between two charges.

Comparing Eqs.\,(\ref{FMC5k}) and (\ref{FMC4k}), we obtain
\begin{equation}
f\left(K_{F}\right)=2\left[\left(K_{F}\right)^{0101}+\left(K_{F}\right)^{0202}\right]\ .\label{fKf}
\end{equation}
This indeed satisfy $f\left(K_{F}\right)\rightarrow f^{\prime}\left(v\right)$
in the particular case defined by Eqs.\,\eqref{paramebire} and \eqref{paramebire2}.
It is therefore reasonable to expect that our results can be generalized
by substituting $f^{\prime}\left(v\right)$ by $f\left(K_{F}\right)$
in the presence of a more generic LV background.

In terms of the LV parameters defined in\,\citep{kosmat1,kosmat2},
we have
\begin{align}
\left(K_{F}\right)^{0101} & =-\frac{1}{2}\left[\left({\tilde{\kappa}}_{e+}\right)^{11}+\left({\tilde{\kappa}}_{e-}\right)^{11}+{\tilde{\kappa}}_{\textrm{tr}}\right]\ ,\label{CLV1}\\
\left(K_{F}\right)^{0202} & =-\frac{1}{2}\left[\left({\tilde{\kappa}}_{e+}\right)^{22}+\left({\tilde{\kappa}}_{e-}\right)^{22}+{\tilde{\kappa}}_{\textrm{tr}}\right]\ ,\label{CLV2}
\end{align}
leading to
\begin{equation}
f\left(K_{F}\right)=f^{\prime}\left(v\right)-\left[\left({\tilde{\kappa}}_{e+}\right)^{11}+\left({\tilde{\kappa}}_{e+}\right)^{22}\right]\ ,\label{fkfp}
\end{equation}
where Eqs.\,\eqref{fv} and \eqref{parameters2} were taken into
account.

Our conclusion is that, in a more general setting, our results will
be modified by a contribution arising from the ${\tilde{\kappa}}_{e+}$
coefficients, as given in Eq.\,\eqref{fkfp}. So, for example, the
interaction energy and force between the charge and the mirror will
be 
\begin{align}
E_{MC}^{\prime} & =E_{MC}+\frac{q^{2}}{16\pi R}\left[\left({\tilde{\kappa}}_{e+}\right)^{11}+\left({\tilde{\kappa}}_{e+}\right)^{22}\right]{\cal {F}}_{2}\left(mR\right)\ ,\label{EMCKe}\\
F_{MC}^{\prime} & =F_{MC}+\frac{q^{2}}{16\pi R^{2}}\left[\left({\tilde{\kappa}}_{e+}\right)^{11}+\left({\tilde{\kappa}}_{e+}\right)^{22}\right]{\cal {F}}_{1}\left(mR\right)\ ,
\end{align}
where $E_{MC}$ and $F_{MC}$ are given by Eqs. \eqref{energyC3}
and \eqref{FC2}, respectively.

It is important to mention, however, that even if the $K_{F}$ tensor
has nineteen independent components, just a subset of those can appear
in the interaction that we are considering in this work.

Next, to obtain the resulting torque in the more general setting,
it is convenient to write the interaction energy explicitly in terms
of $K_{F}$, 
\begin{align}
E'_{MC} & =-\frac{q^{2}}{16\pi R}\Biggl[1-2mRe^{2mR}Ei\left(1,2mR\right)\label{EMCKF12}\\
 & +2\left[\left(K_{F}\right)^{0101}+\left(K_{F}\right)^{0202}\right]{\cal {F}}_{2}\left(mR\right)\Biggr]\ .
\end{align}
Defining a tree-dimensional vector
\begin{equation}
{\bf {K}}_{F}=\left(\left(K_{F}\right)^{0101},\left(K_{F}\right)^{0202},\left(K_{F}\right)^{0303}\right)\ ,\label{vectKF}
\end{equation}
we can write
\begin{align}
\left(K_{F}\right)^{0101} & =\mid{\bf {K}}_{F}\mid\sin\left(\theta\right)\cos\left(\phi\right)\ ,\label{KFespheri1}\\
\left(K_{F}\right)^{0202} & =\mid{\bf {K}}_{F}\mid\sin\left(\theta\right)\sin\left(\phi\right)\ ,\label{KFespheri2}
\end{align}
where $0<\theta<\pi$ and $0<\phi<2\pi$ are the polar and azimuthal
angles in spherical coordinates, respectively ($z$ being the polar
axis). Substitution of Eq. (\ref{KFespheri1}) and (\ref{KFespheri2})
in Eq. (\ref{EMCKF12}) leads to
\begin{align}
E'_{MC}\left(\theta,\phi\right) & =-\frac{q^{2}}{16\pi R}\Biggl[1-2mRe^{2mR}Ei\left(1,2mR\right)\nonumber \\
 & +2\mid{\bf {K}}_{F}\mid\sin\left(\theta\right)\left[\cos\left(\phi\right)+\sin\left(\phi\right)\right]{\cal {F}}_{2}\left(mR\right)\Biggr]\ .\label{EMCKFPT}
\end{align}

The interaction energy in Eq.\,(\ref{EMCKFPT}) generates two kind
of torques on the system, depending on its orientation with regards
to the background LV, one related to the $\theta$ angle, and another
to the $\phi$ angle, i.e.,
\begin{align}
\tau'_{MC}\left(\theta\right) & =-\frac{\partial}{\partial\theta}E'_{MC}\left(\theta,\phi\right)\nonumber \\
 & =\frac{q^{2}}{8\pi R}\mid{\bf {K}}_{F}\mid\cos\left(\theta\right)\left[\cos\left(\phi\right)+\sin\left(\phi\right)\right]{\cal {F}}_{2}\left(mR\right)\ ,
\end{align}
and
\begin{align}
\tau'_{MC}\left(\phi\right) & =-\frac{\partial}{\partial\phi}E'_{MC}\left(\theta,\phi\right)\nonumber \\
 & =\frac{q^{2}}{8\pi R}\mid{\bf {K}}_{F}\mid\sin\left(\theta\right)\left[\cos\left(\phi\right)-\sin\left(\phi\right)\right]{\cal {F}}_{2}\left(mR\right)\ .
\end{align}

In order to make an estimation on these torques, we have to write
$\mid{\bf {K}}_{F}\mid$ in terms of the usual LV coefficients, using
Eqs.\,(\ref{CLV1}) and (\ref{CLV2}), together with
\begin{equation}
\mid{\bf {K}}_{F}\mid=\Bigl\{\left[\left(K_{F}\right)^{0101}\right]^{2}+\left[\left(K_{F}\right)^{0202}\right]^{2}+\left[\left(K_{F}\right)^{0303}\right]^{2}\Bigr\}^{1/2}\label{midKF}
\end{equation}
thus obtaining
\begin{equation}
\left(K_{F}\right)^{0303}=-\frac{1}{2}\left[\left({\tilde{\kappa}}_{e+}\right)^{33}+\left({\tilde{\kappa}}_{e-}\right)^{33}+{\tilde{\kappa}}_{\textrm{tr}}\right]\ .\label{K0303}
\end{equation}
Using the same data we considered in the last section, together with
the upper bounds $\left({\tilde{\kappa}}_{e-}\right)^{ij}\sim4\times10^{-18}$
and $\left({\tilde{\kappa}}_{e+}\right)^{ij}\sim2\times10^{-37}$
from\,\citep{VANR}, we obtain an order of magnitude estimate of
$\tau'_{MC}\sim10^{-40}$Nm for a perfect mirror, and $\tau'_{MC}\sim10^{-45}$Nm
for $m\sim10^{-5}$GeV.

As a final note, we mention that obtaining the results of this section
through an explicit calculation, as done in Sec.\,\ref{III}, would
not be a possible task, since the propagators would be much more involved,
and the resulting integrals would result too complicated to be calculated
analytically. The results we present here, however, are founded on
the explicit results obtained in the previous section for the particular
case, as well as the application of the image method, which has been
extensively discussed in a similar setting in\,\citep{LHCBFABplate}.

\section{\label{IV} Conclusions and final remarks}

In this paper we have studied some aspects of the nonbirefringent
CPT-even gauge sector of the SME near a semi-transparent mirror. We
considered a model where the Lorentz symmetry breaking is caused by
a single background vector $v^{\mu}$ and obtained perturbative results
up to second order in this parameter.

We computed the modified Lorentz violating propagator for the gauge
field due to presence of the mirror and calculated the interaction
energy, as well as the interaction force, between the mirror and a
static point-like charge. In the limiting case of perfect mirrors,
we recovered the interaction found via the image method.

We showed that when the charge is placed in the vicinity of the mirror,
a spontaneous torque emerges on this system due to the orientation
of the mirror with respect to the LV background vector. This torque
is a new effect with no counterpart in Maxwell electrodynamics in
the presence of a semi-transparent mirror. We also showed that the
connection between the massless scalar field in the presence of a
Dirichlet plane and the gauge field in the presence of a perfect conductor
remains when we have the corresponding Lorentz violating terms in
both theories.

Then, we used the image method to generalize our results for a more
general Lorentz violating $K_{F}$ coefficient. We were able to obtain
some estimates for the physical effects that it can induce in the
interaction of a point-like charge and a mirror.

As a possible future study, one interesting point would be the inclusion
of the CPT-odd photon sector of the SME, in the presence of semi-transparent
boundaries \citep{inprep}.

\textbf{\medskip{}
}

\textbf{Acknowledgments.} This study was financed in part by the Coordenação
de Aperfeiçoamento de Pessoal de Nível Superior -- Brasil (CAPES)
-- Finance Code 001 (LHCB), and Conselho Nacional de Desenvolvimento
Científico e Tecnológico (CNPq) via the grant 305967/2020-7 (AFF).


\begin{thebibliography}{99}
\bibitem{SME1} D. Colladay, V. A. Kostelecký, Phys. Rev. D \textbf{{55}},
6760 (1997).

\bibitem{SME2} D. Colladay, V. A. Kostelecký, Phys. Rev. D \textbf{{58}},
116002 (1998).

\bibitem{LHCFABJHN} L.H.C. Borges, F.A. Barone, J. A. Helayel-Neto,
Eur. Phys. J. C \textbf{{74}}, 2937 (2014).

\bibitem{fontes2} L. H. C. Borges, A. F. Ferrari, F. A. Barone, Eur.
Phys. J. C \textbf{{76}}, 599 (2016).

\bibitem{fontes3} L.H.C. Borges, F.A. Barone, Braz. J. Phys. \textbf{{49}},
571-582 (2019).

\bibitem{CED1} H. Belich, M.M. Ferreira, J.A. Helayel-Neto and M.T.D.
Orlando, Phys. Rev. D \textbf{{68}}, 025005 (2003).

\bibitem{CED2} Q.G. Bailey and V.A. Kostelecký, Phys. Rev. D \textbf{{70}},
076006 (2004).

\bibitem{CED3} Rodolfo Casana, Manoel M. Ferreira Jr., Adalto R.
Gomes, and Frederico E. P. dos Santos, Phys. Rev. D \textbf{{82}},
125006 (2010).

\bibitem{CED4} Rodolfo Casana, Manoel M. Ferreira Jr., Adalto R.
Gomes, and Paulo R. D. Pinheiro Phys. Rev. D \textbf{{80}}, 125040
(2009).

\bibitem{CED5} Rodolfo Casana, Manoel M. Ferreira Jr, and Carlos
E. H. Santos, Phys. Rev. D \textbf{{78}}, 105014 (2008).

\bibitem{CED6} L.H.C. Borges, F.A. Barone, A.F. Ferrari, Europhys.
Lett. \textbf{{122}}, 31002 (2018).

\bibitem{CED7} Manoel M. Ferreira Jr., Letí cia Lisboa-Santos, Roberto
V. Maluf, and Marco Schreck, Phys. Rev. D \textbf{{100}}, 055036
(2019).

\bibitem{CED8} Rodolfo Casana, Manoel M. Ferreira Jr., Letícia Lisboa-Santos,
Frederico E. P. dos Santos, Marco Schreck, Phys. Rev. D \textbf{{97}},
115043 (2018).

\bibitem{LHCAFFcptodd} L.H.C. Borges and A.F. Ferrari, Mod.Phys.Lett.A
37 (2022) 04, 2250021.

\bibitem{QED1} F.R. Klinkhamer and M. Schreck, Nuc. Phys. B \textbf{{848}},
90 (2011).

\bibitem{QED2} M.A. Hohensee, R. Lehnert, D.F. Phillips and R.L.
Walsworth, Phys. Rev. D \textbf{{80}}, 036010 (2009).

\bibitem{QED3} D. Colladay and V.A. Kostelecký, Phys. Lett. B \textbf{{511}},
209 (2001).

\bibitem{QED4} B. Charneski, M. Gomes, R.V. Maluf, and A. J. da Silva,
Phys. Rev. D \textbf{{86}}, 045003 (2012).

\bibitem{QED5} G.P. de Brito, J.T. Guaitolini Junior, D. Kroff, P.C.
Malta, and C. Marques, Phys. Rev. D \textbf{{94}}, 056005 (2016).

\bibitem{QED6} Frederico E.P. dos Santos and Manoel M. Ferreira,
Symmetry \textbf{{10}}, 302 (2018).

\bibitem{QED7} J.C.C. Felipe, H.G. Fargnoli, A.P. Baeta Scarpelli,
and L.C.T. Brito, Int. Jour. of Mod. Phys. A \textbf{{34}}, 1950139
(2019).

\bibitem{QED8} A.J.G. Carvalho, A.F. Ferrari, A.M. de Lima, J.R.
Nascimento, A. Yu. Petrov, Nucl. Phys. B \textbf{{942}}, 393-409
(2019).

\bibitem{RC1} R. Jackiw and V.A. Kostelecký, Phys. Rev. Lett. \textbf{{82}},
3572 (1999).

\bibitem{RC2} A.P. Baêta Scarpelli, M. Sampaio, M.C. Nemes, and B.
Hiller, Eur. Phys. J. C \textbf{{56}}, 571 (2008).

\bibitem{RC3} J.R. Nascimento, E. Passos, A.Yu. Petrov, and F.A.
Brito, JHEP \textbf{{0706}}, 016 (2007).

\bibitem{RC4} T. Mariz, J.R. Nascimento, E. Passos, R.F. Ribeiro,
and F.A. Brito, JHEP \textbf{{0510}}, 019 (2005).

\bibitem{RC5} M. Pérez-Victoria, JHEP \textbf{{0104}}, 032 (2001).

\bibitem{RC6} G. Bonneau, Nucl. Phys. B \textbf{{593}}, 398 (2001).

\bibitem{RC7} B. Altschul, Phys. Rev. D \textbf{{99}}, 111701(R)
(2019).

\bibitem{RC8} L.H.C. Borges, A.G. Dias, A.F. Ferrari, J.R. Nascimento,
A. Yu. Petrov, Phys. Rev. D \textbf{{89}}, 045005 (2014).

\bibitem{RC9} A. F. Ferrari, J. R. Nascimento, A. Yu. Petrov, Eur.
Phys. J. C \textbf{{80}}, 459 (2020).

\bibitem{TPD1} A. de Souza Dutra, M. Hott, and F.A. Barone, Phys.
Rev. D \textbf{{74}}, 085030 (2006).

\bibitem{TPD2} M.N. Barreto, D. Bazeia, and R. Menezes, Phys. Rev.
D \textbf{{73}}, 065015 (2006).

\bibitem{TPD3} A. de Souza Dutra, and R.A.C. Correa, Phys. Rev. D
\textbf{{83}}, 105007 (2011).

\bibitem{TPD4} R. Casana, M.M. Ferreira Jr., E. da Hora, and A.B.F.
Neves, Eur. Phys. J. C \textbf{{74}}, 3064 (2014).

\bibitem{wave1} L.H.C. Borges, A.G. Dias, A.F. Ferrari, J.R. Nascimento,
A.Yu. Petrov, Phys. Lett. B \textbf{{756}}, 332 (2016).

\bibitem{wave2} B. Agostini, F.A. Barone, F.E. Barone, Patricio Gaete,
J. A. Helayel-Neto, Phys. Lett. B \textbf{{708}}, 212 (2012).

\bibitem{G1} L.H.C. Borges and D. Dalmazi, Phys. Rev. D \textbf{{99}},
024040 (2019).

\bibitem{G2} R. Jackiw and S. Y. Pi, Phys. Rev. D \textbf{{68}},
104012 (2003).

\bibitem{G3} A. F. Ferrari, M. Gomes, J. R. Nascimento, E. Passos,
A. Yu. Petrov, A. J. da Silva, Phys. Lett. B \textbf{{652}}, 174-180
(2007).

\bibitem{G4} B. Pereira-Dias, C. A. Hernaski and J. A. Helayel-Neto,
Phys. Rev. D \textbf{{83}}, 084011 (2011).

\bibitem{G5} R. V. Maluf, C. A. S. Almeida, R. Casana, M. M. Ferreira
Jr., Phys. Rev. D \textbf{{90}}, 025007 (2014).

\bibitem{G6} V. Alan Kostelecký and Matthew Mewes, Phys. Lett. B
\textbf{{757}}, 510-514 (2016).

\bibitem{G7} V. Alan Kostelecký and Matthew Mewes, Phys. Lett. B
\textbf{{779}}, 136-142 (2018).

\bibitem{NC1} A. Anisimov, T. Banks, M. Dine, M. Graesser, Phys.
Rev. D \textbf{{65}}, 085032 (2002).

\bibitem{NC2} C.E. Carlson, C.D. Carone, R.F. Lebed, Phys. Lett.
B \textbf{{518}}, 201 (2001).

\bibitem{NC3} S. Aghababaei and M. Haghighat Phys. Rev. D \textbf{{96}},
075017 (2017).

\bibitem{Milton} K.A. Milton, \textit{{The Casimir Effect, Physical
Manifestations of Zero-Point Energy, World Scientific, Singapore}}
(2001).

\bibitem{BorUM} M. Bordag, U. Mohideen, and V. M. Mostepanenko, Phys.
Rep. \textbf{{353}}, 1 (2001).

\bibitem{KimballA} Kimball A. Milton, J. Phys. A: Math. Gen.\textbf{{37}},
63916406 (2004).

\bibitem{BordKD} M. Bordag, K. Kirsten and D. Vassilevich, Phys.
Rev. D \textbf{{59}}, 085011 (1999).

\bibitem{NRVMH} N. Graham, R.L. Jaffe, V. Khemani, M. Quandt, M.
Scandurra and H. Weigel, Nucl. Phys. B \textbf{{645}}, 49 (2002).

\bibitem{NRVMMH2} N. Graham, R.L. Jaffe, V. Khemani, M. Quandt, M.
Scandurra and H. Weigel, Phys. Lett. B \textbf{{572}}, 196 (2003).

\bibitem{PsRj} P. Sundberg and R.L. Jaffe, Annals Phys. \textbf{{309}},
442 (2004).

\bibitem{Caval} R.M. Cavalcanti, {[}arXiv:hep-th/0201150{]}.

\bibitem{FABFEB2} F. A. Barone and F. E. Barone, Eur. Phys. J. C
\textbf{{74}}, 3113 (2014).

\bibitem{FABFEB} F.A. Barone and F.E. Barone, Phys. Rev. D \textbf{{89}},
065020 (2014).

\bibitem{GTFABFEB} G. T. Camilo, F. A. Barone and F. E. Barone, Phys.
Rev. D \textbf{{87}}, 025011 (2013).

\bibitem{OliveiraBorgesAFF}H.L. Oliveira, L.H.C. Borges, F.E. Barone
and F.A. Barone, Eur. Phys. J. C 81, 558 (2021).

\bibitem{FABAAN1} F. A. Barone and A. A. Nogueira, Eur. Phys. J.
C \textbf{{75}}, 339 (2015).

\bibitem{LW1} F.A. Barone and A.A. Nogueira, Int. J. Mod. Phys.:
Conf. Ser. \textbf{{41}}, 1660134, (2016).

\bibitem{LW2} M. Blazhyevska, J. of Phys. Stud. \textbf{{16}},
3001 (2012).

\bibitem{LW3} L.H.C. Borges, A.A. Nogueira, E.H. Rodrigues, F.A.
Barone, Eur. Phys. J. C \textbf{{80}}, 1082 (2020).

\bibitem{LHCBFEBHLO} L. H. C. Borges, F. E. Barone, C. C. H. Ribeiro,
H. L. Oliveira, R. L. Fernandes, F. A. Barone, Eur. Phys. J. C \textbf{{80}},
238 (2020).

\bibitem{BorgesBarone22}L.H.C. Borges and F.A. Barone, Phys. Lett.
B 824, 136759 (2022).

\bibitem{CSE1} M.B. Cruz, E.R. Bezerra de Mello and A. Yu. Petrov,
Phys. Rev. D \textbf{{96}}, 045019 (2017).

\bibitem{CSE2} M.B. Cruz, E.R. Bezerra de Mello and A. Yu. Petrov,
Mod. Phys. Lett. A \textbf{{33}}, 1850115 (2018).

\bibitem{CSE3} M. Frank, I. Turan, Phys. Rev. D \textbf{{74}},
033016 (2006).

\bibitem{CSE4} A.F. Santos, F.C. Khanna, Phys. Lett. B \textbf{{762}},
283 (2016).

\bibitem{CSE5} L. M. Silva, H. Belich, J. A. Helayel-Neto, arXiv:1605.02388
(2016).

\bibitem{CSE6} A. Martí n-Ruiz, C.A. Escobar, Phys. Rev. D \textbf{{94}},
076010 (2016).

\bibitem{CSE7} Dêivid R. da Silva and E. R. Bezerra de Mello, arXiv:2006.12924
(2020).

\bibitem{CSE8} A. Martí n-Ruiz, C.A. Escobar, A.M. Escobar-Ruiz,
and O.J. Franca, Phys. Rev. D \textbf{{102}}, 015027 (2020).

\bibitem{CSE9}Amirhosein Mojavezi, Reza Moazzemi, Mohammad Ebrahim
Zomorrodian, Nucl. Phys. B \textbf{{941}}, 145-157 (2019).

\bibitem{CSE10} M.B. Cruz, E.R. Bezerra de Mello, and A. Yu. Petrov,
Phys. Rev. D \textbf{{99}}, 085012 (2019).

\bibitem{CSE11}C. A. Escobar, Leonardo Medel, and A. Martí n-Ruiz,
Phys. Rev. D \textbf{{101}}, 095011 (2020).

\bibitem{CSE12} Andrea Erdas, arXiv:2005.07830 (2020).

\bibitem{CSE13} M.A. Valuyan, Mod. Phys. Lett. A \textbf{{35}},
2050149 (2020).

\bibitem{CSE14} M.B. Cruz, E.R. Bezerra de Mello, and H. F. Santana
Mota, Phys. Rev. D \textbf{{102}}, 045006 (2020).

\bibitem{CSE15} Massimo Blasone, Gaetano Lambiase, Luciano Petruzziello,
and Antonio Stabile, Eur. Phys. J. C \textbf{{78}}, 976 (2018).

\bibitem{LHCBFABplate} L.H.C. Borges and F.A. Barone, Eur. Phys.
J. C \textbf{{77}}, 693 (2017).

\bibitem{LHCBFABplate2} L.H.C. Borges and F.A. Barone, Braz. J. Phys.
\textbf{{50}}, 647-657 (2020).

\bibitem{LHCBAFFFAB} L. H. C. Borges, A. F. Ferrari and F. A. Barone,
Nucl. Phys. B \textbf{{954}}, 114974 (2020).

\bibitem{LHCFAB} L.H.C. Borges and F.A. Barone, Eur. Phys. J. C \textbf{{76}},
64 (2016).

\bibitem{BAlt} B. Altschul, Phys. Rev. Lett. \textbf{{98}}, 041603
(2007).

\bibitem{Gradshteyn} I.S. Gradshteyn and I.M. Ryzhik, \textit{{Table
of Integrals, Series, and Products}}, Academic Press (2000).

\bibitem{Arfken} G.B. Arfken, H.J. Weber, \textit{{Mathematical
Methods for Physicists}} (Academic Press, USA, 1995).

\bibitem{FRKlin} F.R. Klinkhamer and M. Risse, Phys. Rev. D \textbf{{77}},
117901 (2008).

\bibitem{VANR} V.A. Kostelecký, N. Russell. arXiv:0801.0287v15 (2022).

\bibitem{kosmat1}V. A. Kostelecký and M. Mewes, Phys. Rev. Lett.
87, 251304 (2001).

\bibitem{kosmat2}V. A. Kostelecký and M. Mewes, Phys. Rev. D 66,
056005 (2002).

\bibitem{inprep} L. H. C. Borges, A. F. Ferrari, in preparation.
\end{thebibliography}
\end{document}